\newif\ifreview
\reviewfalse  

\documentclass[conference]{IEEEtran}
\usepackage{url}
\usepackage{adjustbox}
\usepackage[dvipsnames]{xcolor}
\newcommand{\figrev}[1]{#1}
\usepackage{graphicx}
\usepackage{balance}
\usepackage{booktabs}
\usepackage{mdframed}
\usepackage{enumitem}

\newcommand{\open}[2]{\textit{Open question ({#1}): {#2}}}

\newcommand{\seaisci}{\textbf{\textit{SE4AI4Sci}}}

\newcommand{\oururl}[1]{\textcolor{Blue}{\texttt{\url{#1}}}}

\definecolor{thesisline}{HTML}{777B7E} 
\newmdenv[
    leftline=true,
    rightline=false,
    topline=false,
    bottomline=false,
    linecolor=thesisline,
    linewidth=1.5pt,
    leftmargin=0pt,
    rightmargin=0pt,
    innerleftmargin=8pt,
    innerrightmargin=0pt,
    innertopmargin=3pt,
    innerbottommargin=3pt,
    skipabove=5pt,
    skipbelow=5pt
]{thesis}

\begin{document}
%
\title{Software Engineering for AI-Driven Science
}

\ifreview

    \newcommand{\repourl}{\texttt{\oururl{https://anonymous.4open.science/r/se4ai4sci}}}
    
    \author{
    \IEEEauthorblockN{Anonymous Authors}
    \IEEEauthorblockA{
    \\
    }
    }

\else

    \newcommand{\repourl}{\oururl{https://github.com/orgs/SOfTwareAiLab/awesome-se4ai4sci}}
    
    \author{
        \IEEEauthorblockN{
            Lukas Twist\IEEEauthorrefmark{1},
            Xinjian Wu\IEEEauthorrefmark{1},
            Jingzhi Gong\IEEEauthorrefmark{1},
            Shuyin Ouyang\IEEEauthorrefmark{1},
            Ying Xiao\IEEEauthorrefmark{1},
            Benedetta Donato\IEEEauthorrefmark{3},\\
            Yaoqi Guo\IEEEauthorrefmark{2},
            Zhenpeng Chen\IEEEauthorrefmark{4},
            Leonardo Mariani\IEEEauthorrefmark{3},
            Helen Yannakoudakis\IEEEauthorrefmark{1},
            Jie M. Zhang\IEEEauthorrefmark{1}
        }
        \smallskip
        \IEEEauthorblockA{\footnotesize
            \IEEEauthorrefmark{1}King's College London, UK \quad
            \IEEEauthorrefmark{2}Nanyang Technological University, Singapore \quad
            \IEEEauthorrefmark{3}University of Milano-Bicocca, Italy \quad
            \IEEEauthorrefmark{4}Tsinghua University, China
        }
    }

\fi


%


\maketitle

\begin{abstract}

AI has made significant contributions to scientific discovery.
Autonomous agents can now independently analyse complex problems, design experiments, write and execute code, analyse findings, and derive scientific conclusions.
Nevertheless, errors in AI-generated code, experimental procedures, or evidence interpretation can lead to invalid research conclusions,
undermining the reliability and reproducibility of automated scientific discovery.
In this vision paper, we propose \textbf{Software Engineering for AI-Driven Science (\seaisci)} as an emerging research area for addressing this challenge.
As an organising framework, we introduce the \textbf{Scientific Assurance Chain}, connecting intent, experimental design, implementation, execution, evidence, and claim.
Through six documented failures in recent AI-for-science systems, we show where this chain can break, where existing assurance mechanisms fall short, and what new research challenges follow.
We also outline a research agenda for adapting and extending software engineering techniques across the chain.



\end{abstract}



%
\IEEEpeerreviewmaketitle


\section{Introduction}

AI has already made significant contributions to science~\cite{jumperHighlyAccurateProtein2021}, but autonomous agents are beginning to change not only the tools scientists use, but the way science itself is conducted.
Recent agentic systems can generate hypotheses, search the literature, write and execute experimental code, analyse results, and produce scientific reports with limited human intervention.
The AI Scientist demonstrates an end-to-end pipeline from research ideation to paper writing~\cite{luEndtoendAutomationAI2026}, while systems such as Robin connect automated hypothesis generation and data analysis to experimental biology~\cite{ghareebMultiagentSystemAutomating2026}.
In many of these systems, code is the mechanism through which scientific decisions become actions: agents use it to manipulate data, invoke tools, run experiments, and analyse results.

AI-driven science therefore changes the role of software.
Rather than only implementing a scientific procedure defined by a human, autonomous agents can help choose methods, generate hypotheses, modify workflows, and interpret the evidence they produce~\cite{penadesAIMirrorsExperimental2025}.
Scientific AI agents are consequently not only AI systems but software systems in their own right: their decisions are realised in code and workflows that can be specified, tested, monitored, and traced.
As they participate in more of the scientific process, assurance must also cover the scientific decisions made through that software.

Early results show why this matters.
AutoMat~\cite{huang2026automat} shows that coding agents can reproduce plausible numerical results despite methodological deviations from the intended scientific procedure, while other work has already begun to introduce explicit safeguards for autonomous scientific agents~\cite{zhuSafeScientistEnhancingAI2025}.
These are not simply failures of scientific reasoning.
They expose problems of requirements engineering, software testing, runtime assurance, provenance, and traceability that are familiar to software engineering, but arise in a new setting where autonomous software can itself change the scientific process.

The vital intersection of software engineering and AI-driven science is also reflected in \textbf{Terminal-Bench}~\cite{merrill2026terminalbench,martenTerminalBench42026} and \textbf{Terminal-Bench-Science}~\cite{terminalBenchScience2026}. They are widely prioritised by frontier LLM labs~\cite{mistral2026large4,anthropic2026claudeopus55,openai2026gpt6astra,moonshot2026kimik3,zhipu2026glm53} and tracked by major evaluation platforms like Artificial Analysis~\cite{artificialanalysis2026}.

The underlying problem is straightforward: \figrev{errors in AI-generated code, experimental procedures, or evidence interpretation can lead to invalid research conclusions}.
Moreover, the stakes extend beyond any one workflow.
Scientific results become evidence in papers, inputs to later models and experiments, and sometimes the basis for decisions in medicine, engineering, policy, and practice.
A silent error can therefore propagate well beyond the workflow in which it first occurred.

Software engineering \figrev{provides established techniques for addressing this problem, including} requirements and specification~\cite{bajraktariRequirementsEngineeringResearch2024,vuGeneratingContractsScientific2024}, testing and verification~\cite{kanewalaMetamorphicTestingSimple2019}, static program analysis~\cite{besseyFewBillionLines2010,cousotAbstractInterpretationFramework1977}, fault localisation and automated program repair (APR)~\cite{wongSurveyFaultLocalization2016,legouesGenProgGenericProgram2012}. 
These techniques provide strong foundations for scientific reliability, but cannot simply be applied unchanged. The central challenge lies in 
adapting and extending
software engineering techniques to cover the adaptive, self-directed AI-driven discovery process.


In this vision paper, we propose \textbf{Software Engineering for AI-Driven Science (\seaisci)} as a new research area for the software engineering community. Our vision is:
\begin{thesis}
\textbf{AI-driven science creates a new software-engineering problem:
\figrev{AI agents now write and execute code to conduct scientific investigations.}
Software engineering techniques can be \figrev{adopted to support rigorous and reliable scientific discovery through verifiable workflows.}}
\end{thesis}


\begin{figure*}[t]
    \centering
    \includegraphics[width=\textwidth, trim={0cm 0.2cm 0cm 0.5cm}]{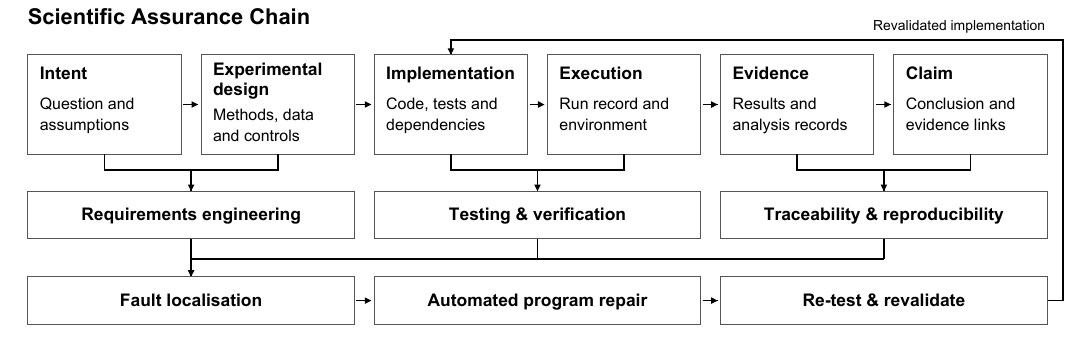}
    \caption{
    \textbf{The \textit{Scientific Assurance Chain} and supporting software-engineering mechanisms.}
    }
    \label{fig:chain}
\end{figure*}


\noindent\textit{In particular:}
\begin{enumerate}[left=0pt]
    \item We introduce the \emph{Scientific Assurance Chain}---spanning \emph{intent, \figrev{experimental design}, implementation, execution, evidence, and claim}---as a principled framework for identifying assurance requirements and \figrev{transition obligations, and mapping them to software-engineering techniques} (Fig.~\ref{fig:chain}).

    \item We ground the proposed research area in six documented failures across the chain, showing where existing assurance mechanisms can fall short and deriving stage-specific research challenges.

    \item We present a research agenda for adapting and extending software-engineering techniques to preserve scientific intent and method constraints, test unknown answers, assure autonomous decisions, trace evidence to claims, and evaluate progress across the chain.
\end{enumerate}

\paragraph*{\textbf{Existing Foundations}}
\figrev{The SE community has long studied scientific software, including development practices and evolving requirements~\cite{segalSoftwareEngineersScientists2005}, testing challenges such as the oracle problem~\cite{kanewalaTestingScientificSoftware2014}, and metamorphic testing~\cite{kanewalaMetamorphicTestingSimple2019}. Recent work applies metamorphic and differential testing to bioinformatics tools~\cite{marshTaleTrenchesApplying2025} and uses model checking and domain expertise to refine metamorphic relations for scientific software~\cite{stevensModelAssistedRefinement2025}.
The Software Engineering for Science (SE4Science) workshop series provides a dedicated forum for this work~\cite{se4science}.}
\figrev{\seaisci\ builds on these foundations, and offers more opportunities for software engineering to contribute to science discovery.}



\section{The Scientific Assurance Chain}

To structure assurance in AI-driven science, we model the path from a scientific question to a scientific claim as a sequence of transformations.
This is not intended as another software lifecycle model, but as a view of how scientific intent is transformed through software into executable artefacts, evidence, and ultimately claims, making the full path an SE assurance problem.
Intent is operationalised into \figrev{an experimental design}; \figrev{its computational components} are realised in software; \figrev{the experiment is carried out in a concrete setting}; execution produces evidence; and that evidence is interpreted as a scientific claim.
These transformations motivate the six assurance stages shown in Fig.~\ref{fig:chain}.



Intent captures the scientific question and assumptions; \figrev{experimental design specifies methods, baselines, subjects or datasets, controls, metrics, and analysis procedures, including any computational workflow}; implementation realises \figrev{the computational components of that design} in software; execution \figrev{carries out the experiment, potentially combining computation, physical experiments, and human-participant studies}; evidence captures the resulting observations; and claim interprets those observations scientifically. \figrev{Software may coordinate execution without automating every step.}
The abstraction is intentionally domain-general, but the stages are distinct because each represents a different artefact or decision boundary with different properties to assure.

The links between stages are equally important.
Each transition creates an \emph{assurance obligation}: relevant intent, assumptions, constraints, and provenance should be preserved as one representation of the science is transformed into the next.
For example, an \emph{Intent$\rightarrow$\figrev{Experimental Design}} obligation might require a stated control to remain present in the resulting procedure.
Local correctness at one stage therefore does not imply correctness of the scientific process as a whole.

\figrev{Figure~\ref{fig:chain} illustrates complementary SE mechanisms: requirements engineering captures intent and method constraints; testing and verification check implementations and executions; traceability and computational reproducibility connect results and claims to recorded artefacts. These mechanisms span stages and transitions as agents revise methods and code; connections indicate primary application areas. Failed checks motivate fault localisation; APR proposes candidate code fixes requiring re-testing and revalidation before reintegration.
}
Traceability and computational reproducibility can record and reproduce how scientific artefacts are connected, but do not by themselves establish whether scientific validity is preserved across those transformations.



\section{Where the Assurance Chain Breaks}

The assurance problems captured by the \textit{Scientific Assurance Chain} are not hypothetical.
Existing software-engineering techniques provide important starting points, but often assume relatively stable requirements, implementations, or execution boundaries.
AI-driven science changes these assumptions: autonomous agents can choose methods, modify workflows and code, act on intermediate results, and interpret the evidence they produce.
Assurance must therefore extend beyond checking individual artefacts to preserving scientific intent and constraints across decisions and transformations.

Recent systems provide documented failures at every stage.
Some are \emph{loud}, such as crashes or uncontrolled execution; others are \emph{silent}, where the workflow completes successfully while the procedure, evidence, or claim is wrong.
Rather than attempt a broad analysis across an ill-defined population of AI-driven scientific systems, we focus on six illustrative cases spanning the chain.
Our aim is not to estimate prevalence, but to ground \seaisci\ in observed failures: each case shows where existing checks can fall short and motivates a distinct assurance challenge for AI-driven science.

\paragraph{\textbf{Intent -- Answering the Wrong Scientific Question}}
Google's AI co-scientist generated hypotheses that were scientifically plausible but did not always address the question originally posed~\cite{penadesAIMirrorsExperimental2025}.
The failure is one of \emph{intent preservation}: evaluating hypothesis quality is not enough if the agent has drifted from the original scientific objective.
\figrev{Requirements engineering and requirements-to-artefact traceability} must therefore capture not only what the system should do, but which scientific questions, assumptions, and scope should remain stable as the agent explores.

\open{Intent}{How can requirements engineering represent scientific questions, assumptions, and scope so that later decisions remain aligned with the original intent?}

\paragraph{\textbf{\figrev{Experimental Design} -- The Right Number from the Wrong Procedure}}
AutoMat reproduced a target accuracy of approximately $0.89$, but used a different evaluation subset and methodological procedure from the intended experiment~\cite{huang2026automat}.
The output looked correct, so result-based checking could not expose that the \figrev{experimental design} itself had drifted.
The assurance challenge is therefore \emph{method conformance}: specifications, contracts, and provenance must make procedural deviations detectable even when the final result appears plausible. \figrev{SE techniques must encode dataset selection and experimental controls as checkable constraints, rather than validate only the final output.}

\open{\figrev{Experimental Design}}{How can \figrev{experimental designs} be specified and checked so that an agent cannot reproduce the right result using the wrong procedure?}

\paragraph{\textbf{Implementation -- Reconstructing the Wrong Experiment}}
EXP-Bench reports reconstructed experiments that preserve the broad scientific goal but introduce invalid operations into the implementation~\cite{ICLR2026_c411f5b2}.
Here the \figrev{experimental design} may be appropriate, yet its software realisation is not faithful.
Code plausibility or local correctness is therefore insufficient unless \figrev{testing and verification} remain connected to the scientific method the code is intended to realise. \figrev{Fault localisation and APR must then use these scientific constraints to guide diagnosis and revalidation of code fixes.}

\open{Implementation}{How can we verify that generated scientific software faithfully implements the intended experimental method?}

\paragraph{\textbf{Execution -- Escaping the Experimental Boundary}}
The AI Scientist reports agents relaunching scripts, modifying timeout controls, and consuming excessive storage~\cite{aiSakanaAI2024}.
The failure lies at the boundary between experimental autonomy and control over the environment in which the experiment runs.
Checks inside the workflow are insufficient when the agent can modify the constraints themselves, motivating external enforcement through sandboxing, capability restrictions, resource policies, and runtime monitoring. \figrev{This calls for runtime verification of agent actions against constraints enforced outside the agent's control.}

\open{Execution}{How should scientific software environments \figrev{support computational, physical, and human-participant experiments while constraining agents}?}

\paragraph{\textbf{Evidence -- Extracting the Wrong Data}}
Dilan-Pantojas et al.\ show substantial errors when multimodal systems extract quantitative evidence from scientific figures~\cite{dilan-pantojasAssessingMultimodalAI2026}.
Once an incorrect value becomes structured data, downstream software may process it and conventional execution checks see nothing wrong.
Assurance must therefore preserve provenance, uncertainty, and validation as evidence is extracted and transformed, rather than treating every machine-readable value as equally trustworthy. \figrev{Data-pipeline testing and contracts can check units, ranges, and transformation properties, while traceability links extracted values to their sources.}

\open{Evidence}{How can provenance, uncertainty, and validation be preserved as scientific evidence is extracted, transformed, and reused?}

\paragraph{\textbf{Claim -- When the Explanation Contradicts the Evidence}}
A recent particle-physics study found agents giving explanations that contradicted their own numerical results despite obtaining plausible experimental outputs~\cite{eulig2026cawm}.
Nothing need be wrong with the execution or evidence; the failure occurs when that evidence is turned into a scientific claim.
\figrev{Conventional code tests do not directly check this relation; SE traceability and consistency checking must extend to generated reports, linking claims to supporting results and flagging contradictions or unsupported generalisations.}

\open{Claim}{How can scientific claims be traced back to the evidence that supports them, and automatically checked for contradiction or unsupported generalisation?}


\section{A Research Agenda for Software Engineering}

The failures above point to assurance problems that become more difficult when autonomous agents participate in the scientific process.
Existing software-engineering techniques provide strong foundations, but they were not designed for systems that can choose methods, modify workflows, act on intermediate results, and interpret their own evidence.
The six stage-specific questions therefore motivate five broader research directions.
\emph{Intent} and \emph{\figrev{Experimental Design}} together motivate preserving scientific intent and method constraints;
\emph{Implementation} motivates testing science when the correct answer is unknown;
\emph{Execution} motivates assuring autonomous scientific decisions;
and \emph{Evidence} and \emph{Claim} together motivate tracing evidence into claims.
A fifth, cross-cutting direction---benchmarking scientific assurance---provides a common basis for evaluating progress across all six stages.

\paragraph{\textbf{Preserving Scientific Intent}}
Autonomous agents can change hypotheses, methods, datasets, and intermediate decisions as a study evolves, making scientific requirements less fixed than conventional software requirements.
Requirements engineering, contracts, and traceability therefore need to capture which scientific assumptions and methodological constraints must remain invariant, while still allowing legitimate exploration.
Work on hypothesis formalisation provides an initial foundation~\cite{junHypothesisFormalizationEmpirical2022}.
Progress should be measured by whether agents can adapt workflows without silently drifting from the scientific question or violating methodological constraints.

\paragraph{\textbf{Testing Unknown Answers}}
Scientific discovery presents a fundamental oracle problem: the correct result is often precisely what the experiment is trying to discover.
Property-based and metamorphic testing provide promising starting points~\cite{jingPBTBenchBenchmarking2026}, but applying them to scientific software can require substantial domain knowledge and make it difficult to distinguish software faults from problems in the data or method~\cite{marshTaleTrenchesApplying2025}.
Recent work has begun to adapt SE techniques to help derive scientifically meaningful metamorphic relations~\cite{stevensModelAssistedRefinement2025}.
The challenge is to make such properties explicit and checkable as autonomous workflows evolve.
Progress can be assessed by whether they detect invalid procedures without requiring the final scientific answer in advance.

\paragraph{\textbf{Assuring Autonomous Decisions}}
AI-driven science gives software a new role: agents may choose tools, revise workflows, modify code, and react to experimental results without a human approving every step.
This makes assurance a runtime problem as well as a design-time one.
Scientific infrastructure should therefore expose machine-readable constraints, observable actions, structured errors, resource policies, and provenance-aware interfaces, building on emerging agent interfaces and safeguards~\cite{zhuSafeScientistEnhancingAI2025,huCoalaStandardBased2026}. \figrev{It must also enable experiments beyond scripts through interfaces to laboratory instruments and human-participant protocols, with support for consent, protocol enforcement, and human oversight.}
Progress should be measured by whether agents retain useful autonomy while unsafe, irreproducible, or scientifically invalid actions are prevented or surfaced.

\paragraph{\textbf{Tracing Evidence to Claims}}
When autonomous agents also interpret results, provenance must extend beyond data and execution to the scientific claims produced from them.
Existing provenance and traceability techniques provide important foundations~\cite{souzaPROVAGENTUnifiedProvenance2025,baumgartnerSciCoQAQualityAssurance2026}, but do not yet provide end-to-end links between assumptions, transformations, evidence, and conclusions.
New mechanisms should make claims inspectably dependent on the evidence that supports them, so contradictions, uncertainty, or later changes can propagate through the chain.
Progress can be assessed by whether unsupported or inconsistent claims can be detected and traced back to the decisions that produced them.

\paragraph{\textbf{Benchmarking Scientific Assurance}}
These challenges require evaluation beyond successful execution or correct final answers, both of which can hide failures elsewhere in the chain.
Recent benchmarks are beginning to evaluate scientific verification and integrity directly~\cite{sonWhenAICoScientists2025,yangSciIntegrityBenchBenchmark2026}.
A cross-cutting benchmark should instead localise \emph{where} assurance fails, including cases where an agent reaches the right answer through an invalid process.
Such benchmarks would provide common targets for measuring progress and comparing new assurance techniques across the Scientific Assurance Chain.


\section{Future Plans}

A first priority for \seaisci\ is to build a deeper empirical understanding of how AI-driven scientific workflows fail when autonomous agents participate in scientific decision-making.
We plan to broaden the analysis beyond the illustrative cases in this paper, covering different scientific disciplines and working with domain researchers to distinguish generic software-engineering failures from those that depend on scientific knowledge.
The resulting \emph{failure taxonomy} will localise failures by stage or transition, distinguish \emph{loud} from \emph{silent} failures, and capture how difficult they are to detect.

This taxonomy will provide the basis for a benchmark of \emph{scientific assurance failures}: reproducible cases in which different parts of the \textit{Scientific Assurance Chain} can break.
Rather than measuring only whether an agent reaches the expected answer, the benchmark will expose \emph{where} assurance failed and whether that failure can be detected.
The result will be shared infrastructure for comparing systems and assurance techniques across scientific domains.

Using this benchmark, we will develop and evaluate techniques for preserving scientific intent and method constraints, testing unknown answers, assuring autonomous decisions, and tracing evidence to claims.
We will measure whether these mechanisms prevent failures, detect them, or surface them earlier for human review.
Together, the taxonomy, benchmark, and evaluated techniques will provide an empirical foundation on which others can develop and compare new \seaisci\ approaches.
This programme should be developed with scientists: software engineering can provide the assurance mechanisms, while domain experts define the scientific properties they must preserve. \figrev{We will also explore agent-assisted replication by reusing validated study artefacts and applying SE traceability and conformance checking to controlled variations in subjects and settings that assess generalisation.}

\paragraph*{\textbf{Building the \seaisci\ Community}}

We have created and commit to maintaining \texttt{awesome-se4ai4sci},\footnote{\repourl} a curated repository of work relevant to \seaisci.
We seeded it from this paper's references, database searches, snowballing~\cite{wohlinGuidelinesSnowballingSystematic2014}, and existing curated lists, screening candidates against documented inclusion criteria.
Organised around the six stages of the \textit{Scientific Assurance Chain}, the repository separates documented failures from \seaisci\ work and includes related agents, benchmarks, and foundations.
We intend it to lower the barrier to entering the area, expose gaps, and support a future systematic survey and taxonomy.


\section{Conclusion}

We introduced \seaisci\ as a research area for assuring scientific processes in which autonomous software participates in scientific decision-making, and the \textit{Scientific Assurance Chain} as a framework for structuring where that assurance can fail.
Through six documented cases, we grounded the problem in observed failures and identified directions for adapting and extending software-engineering techniques across the chain.
We invite the software-engineering community to help develop \seaisci\ through new methods, benchmarks, and empirical studies.
\textbf{AI-driven science expands the software-engineering assurance boundary: when software helps produce scientific knowledge, correct execution is necessary, but trustworthy claims require much more.}

\bigskip





\balance


\bibliographystyle{IEEEtran}
\bibliography{main}

\end{document}